\documentclass[11pt,a4paper]{article}
\pdfoutput=1

\usepackage{amsmath}
\usepackage{amsfonts}
\usepackage{amsthm}
\usepackage{amssymb}
\usepackage{amscd}
\usepackage[british]{babel}
\usepackage{graphicx}
\usepackage{psfrag}
\usepackage{epsfig}
\usepackage{rotating}
\usepackage{times}

\theoremstyle{plain}

\newcommand{\boxend}{\flushright{$\Box$}}

\newcommand{\N}{{\mathbb N}}               % for natural numbers
\newcommand{\Z}{{\mathbb Z}}               % for entire numbers
\newcommand{\R}{{\mathbb R}}               % for real numbers
\renewcommand{\tilde}{\widetilde}

\begin{document}

\title{Cosmological perturbations in teleparallel Loop Quantum Cosmology}

\author{Jaime Haro$^{a,}$\footnote{E-mail: jaime.haro@upc.edu}
}

\maketitle

{$^a$Departament de Matem\`atica Aplicada I, Universitat
Polit\`ecnica de Catalunya, Diagonal 647, 08028 Barcelona, Spain \\
}

\thispagestyle{empty}

\begin{abstract} Cosmological perturbations in Loop Quantum Cosmology (LQC) are usually studied incorporating either
holonomy corrections,  where the Ashtekar connection is replaced by a suitable sinus function
in order to have a well-defined quantum analogue, or inverse-volume corrections coming from the eigenvalues of the inverse-volume operator.

In this paper we will develop
an alternative
approach to calculate cosmological perturbations in LQC based on the fact that,  holonomy corrected LQC
in the flat Friedmann-Lema{\^\i}tre-Robertson-Walker (FLRW) geometry could be also obtained as
 a particular case of teleparallel $F(T)$ gravity
 (teleparallel LQC). The main idea of our approach is to mix the
simple bounce provided by holonomy corrections in LQC with the non-singular perturbation equations
given by  $F(T)$ gravity, in order to obtain a matter bounce scenario as a viable alternative to slow-roll inflation.

In our study,
we have obtained an scale invariant power spectrum of cosmological perturbations.
 However, the
ratio of tensor to scalar perturbations is of order $1$, which does not agree with the current observations. For this
 reason, we suggest a model where a transition
from the matter domination to a quasi de Sitter phase is produced
 in order to enhance the scalar power spectrum.
\end{abstract}

\vspace{0.5cm}

{\bf Pacs numbers:} 04.50.Kd, 98.80.Bp, 98.80.Qc

%\vspace{0.5cm}

\section{Introduction}

%\vspace{0.5cm}

Loop Quantum Cosmology  in the flat Friedmann-Lema{\^\i}tre-Robertson-Walker
 geometry  containing only holonomy corrections
(not inverse-volume effects) can be built in two different ways. The first one consists in
replacing in the classical Hamiltonian the Ashtekar connection which does not have a well defined quantum operator (see for example \cite{as}),
namely ${\bar c}$,
by the function $\sin(\bar{\mu}{\bar c})/\bar{\mu}$ (holonomy corrected LQC). Then, with the new
holonomy corrected Hamiltonian one obtains the modified Friedmann equation (an ellipse in the plane $(H,\rho)$). The alternative approach  consists in looking
for a  teleparallel $F(T)$-Lagrangian density (teleparallel LQC), where $T=-6H^2$ is the scalar torsion, which leads to the same modified Friedmann
 equation \cite{aho}.

When one deals with cosmological perturbations both formulations could, in principle,  lead to different results. If one considers teleparallel
 LQC, one only has to use the well-known  perturbation equations in $F(T)$ gravity \cite{cdds,ccdds,zh}. On the other hand,
cosmological perturbations in holonomy corrected LQC are performed in the Hamiltonian framework. The idea is very simple, starting from  the perturbed classical
Hamiltonian the way to introduce holonomy corrections, like in isotropic models, is based in the replacement
 ${\bar c}\rightarrow \frac{\sin(n\bar{\mu}{\bar c})}{n\bar{\mu}}$ where $n\in \N\setminus \{0\}$ \cite{bh}.
The problem with this prescription is that the algebra of constrains ceases to be preserved, i.e., the Poisson brackets of
the constrains include additional terms called anomalies. These anomalies can be removed,
and the algebra of constrains restored,
inserting some counter-terms in the holonomy modified
Hamiltonian \cite{cmbg}.
However, some of these counter-terms must contain the Ashtekar connection which does not have a quantum analogue, meaning
that, in principle, it is impossible in the context of LQC to quantify this anomaly-free holonomy modified Hamiltonian.

What is important in both formulations is that they provide  a simple bouncing scenario, which for a universe that is matter-dominated  at early times, could
be an alternative to the inflationary
paradigm (see \cite{b12} for a recent review about the problems related with slow-roll inflation and the alternative bouncing scenarios). Note that, if one only considers inverse-volume corrections,
when the universe is filled by  a field under the action of a non-negative potential (to guarantee a  positive energy density),
 one will obtain a
 non bouncing universe because the Hubble parameter never vanishes
(see equations $(5)$ and $(8)$ of  \cite{bojo}). In that case, there will be a super-inflationary phase at early times that could
solve the horizon and flatness problems that appear in Einstein Cosmology (EC), but to obtain an scale invariant spectrum of
primordial perturbations a quasi de Sitter phase is needed. That is the reason why
authors that
only take into account inverse-volume corrections,
 have to impose an slow-roll phase at early times (see for example
\cite{bct,gbg}). In our work, since we are only interested in models without an slow-roll epoch in the expanding phase, we
will disregard these inverse-volume effects.

 Then the following questions arise: Could holonomy corrected or teleparallel
LQC be a viable alternative to inflationary cosmology? Do all the theoretical results obtained from these formulations of LQC (scale invariance
of scalar and tensor
power spectrum of perturbations, ratio of tensor to scalar perturbations,...) match correctly  the current observational results?

To give reasonable answers to those questions, first at all
 we will show  that  for scalar perturbations both formulations give equations that only differ in the velocity of the sound.
This difference is very important at high energies because,  as has been showed in \cite{cmbg}, in the super-inflationary phase $\dot{H}>0$,
the
square of the velocity of sound in holonomy corrected LQC becomes negative. As a consequence
%the evolution of perturbations breaks-down because the
the equation of evolution changes from hyperbolic to an elliptic one. This phenomenon never happens in teleparallel LQC  where the square of
the velocity
of sound is always positive, giving rise to a hyperbolic equation of evolution for all time.
On the other hand, for tensor perturbations, we find an equation  which is completely different from the one obtained in
holonomy corrected LQC.

Once we have obtained our equations for scalar and tensor perturbations, we get
the corresponding gauge invariant Mukhanov-Sasaki equations, showing  that for a matter bouncing universe both
 power spectra are scale invariant. Moreover, we also show that the ratio of tensor to scalar perturbations, contrary to the ambiguous value
 given by holonomy corrected LQC and the small value found in slow-roll inflation,
 is of the order $1$, as in standard matter bounce $F(T)$ gravity. However, this value is greater than the current observational bound,
which means that in this theory  a mechanism has to be introduced in order to amplify the scalar perturbations.

At the end of the paper we discuss such a mechanism to achieve this bound. It consists in introducing, in the contracting phase,  a transition from
 the matter-domination
to a quasi de Sitter stage. At that stage where  scalar perturbations are enhanced and the ratio of tensor to scalar perturbations will decrease
enough to achieve the desired bound.

The units used in this paper are: $\hbar=c=8\pi G=1$. The following notation is also used:

${}^.=\frac{d}{dt}$ means the derivative with respect to the cosmic time $t$.

${}'=\frac{d}{d\eta}$ means the derivative with respect to the conformal time $\eta$.

\section{LQC in the flat Friedmann-Lema{\^\i}tre-Robertson-Walker geometry}

In this Section we review the way to built LQC from holonomy corrections, and how the dynamical effective equations (Friedmann and Raychauduri)
are deduced in the flat  FLRW geometry. The key point is that the effective Friedmann equation depicts an ellipse in the plane $(H,\rho)$ where
$H$ is the Hubble parameter
and $\rho$ is the energy density. Then, since
in $F(T)$ gravity the corresponding modified Friedmann equation depicts a curve in the plane $(H,\rho)$, what we have done is to find a particular
 $F(T)$ theory whose modified Friedmann equation yields the sought  ellipse.  This shows that LQC containing only holonomy
corrections in the flat FLRW geometry
could be understood as a particular case of teleparallelism.

\subsection{Holonomy corrected LQC in the flat FLRW space-time}

Loop Quantum Cosmology is based in the discrete nature of the space time.
Using the canonical conjugate variables $\bar{p}=a^2$ and the Ashtekar connection  ${\bar c}=\gamma\bar{k}$ (on the classical version  $\bar{k}=\dot{a}$),
where $\gamma
\cong 0.2375$
is the Barbero-Immirzi parameter,
one can build
 the quantum theory choosing as a Hilbert space the
quotient
space of the Besicovitch
space of almost periodic functions
by its subspace of null functions.

Note that the Besicovitch space  is the closure of trigonometric polynomials under
the semi-norm (in the $\bar{c}$-representation)
\begin{equation}\label{1}
 ||\Psi||^2=\lim_{L\rightarrow \infty}\frac{1}{2L}\int_{-L}^{L}|\Psi(\bar{c})|^2 d\bar{c}.
\end{equation}

In fact, all the element of this space have the expansion
$$\Psi({\bar{c}})=\sum_{n\in\Z}\alpha_n
|\mu_n\rangle\equiv\sum_{n\in\Z}\alpha_ne^{i\mu_n{\bar{c}}/2},$$
with $\mu_n\in\R$ and $\alpha_n\in {\mathfrak l}^2$ (the space of
square-summable sequences).

Thus,
in this space,
since the Poisson bracket of these canonically conjugate variables is
$\{\bar{c},
\bar{p}\}=\frac{\gamma}{3}$,
 on can define the operator $\hat{\bar{p}}$ as $\hat{{\bar{p}}}=-\frac{i\gamma}{3}\frac{d}{d{\bar c}}$,
but the  operator $\hat{ \bar c}$ defined by $\hat{\bar c}\Psi({ \bar c})={\bar c}\Psi({\bar c})$ does not belong in
this Hilbert space because it has infinite norm (see for instance \cite{as,h}).
As a consequence,  it is impossible to quantize the gravitational part of the Hamiltonian
\begin{eqnarray}\label{2}
{\mathcal H}^{(0)}_{G}=-3H^2a^3= -\frac{3}{\gamma^2}{ \bar c}^2\sqrt{\bar{p}},\end{eqnarray}
because it contains $\bar c$. To solve  this problem  one can  re-define the Hamiltonian introducing almost periodic functions
that approximate ${\bar c}^2$ for small values of $\bar c$ (holonomy corrections).
This can be done using
the general formulae of Loop Quantum Gravity (LQG) \cite{abl03,t01,aps06a}
\begin{eqnarray}
%&&
%\hspace*{-5mm}
{\mathcal H}_{G,LQC}^{(0)}\equiv-\frac{2}{\gamma^3\bar{\mu}^3}
\sum_{i,j,k}\varepsilon^{ijk} Tr[
h_i(\bar{\mu})h_j(\bar{\mu})h_i^{-1} (\bar{\mu})
% \nonumber \\ &&  \times
h_j^{-1}(\bar{\mu})h_k(\bar{\mu})
\{h_k^{-1}(\bar{\mu}),\bar{p}^{3/2}\}].
%=-\frac{\gamma^2c^2}{2\iota^2}a\sin^2\frac{\mu x}{c},
\end{eqnarray}

In this formula the holonomies are given by
\begin{eqnarray}
h_j(\bar{\mu})\equiv e^{-i\frac{\bar{\mu} x}{2}\sigma_j} =\cos\left(\frac{\bar{\mu}
x}{2}\right)-i\sigma_j\sin\left(\frac{\bar{\mu} x}{2}\right),
\end{eqnarray}
where  the Pauli's matrices $\sigma_j$  have been used.

A simple calculation
gives rise to
the following holonomy modified gravitational Hamiltonian \cite{he,b,dmp}
\begin{equation}\label{3}
 {\mathcal H}_{G,LQC}^{(0)}=-3\sqrt{\bar{p}}\frac{\sin^2(\bar{\mu}{\bar c})}{\bar{\mu}^2\gamma^2 },
\end{equation}
which
using the so-called $\bar{\mu}$-scheme, defined by
 $\bar{\mu}=\lambda/\sqrt{\bar{p}}$,
where $\lambda=\sqrt{\frac{\sqrt{3}}{2}\gamma}$ is the square root of the minimum eigenvalue of the area operator in LQG
 (see for instance \cite{as}),  becomes
\begin{eqnarray}\label{hamiltonian}
 {\mathcal H}_{G,LQC}^{(0)}=-3V\frac{\sin^2(\lambda\beta)}{\lambda^2\gamma^2 },
\end{eqnarray}
where we have introduced the volume $V=a^3$ and its canonically conjugate variable $\beta={\bar c}/\sqrt{\bar{p}}$.

Then, from the Hamilton equation $\dot{V}=\{V,{\mathcal H}_{G,LQC}^{(0)}\}$ one obtains the relation
 $H=\frac{\sin(2\lambda\beta)}{2\lambda\gamma }$,
that together with the Hamiltonian constrain ${\mathcal H}_{G,LQC}^{(0)}+V\rho=0$, lead to the
  effective (or holonomy modified) Friedmann equation in LQC, which
depicts
 the following ellipse
\begin{eqnarray}\label{7}
H^2=\frac{\rho}{3}\left(1-\frac{\rho}{\rho_c}  \right),
\end{eqnarray}
in the plane $(H,\rho)$. In  formula (\ref{7}) $\rho_c=\frac{3}{\lambda^2\gamma^2}\cong 0.4\rho_{{pl}}$ is the so-called {\it critical density}.

\subsection{Teleparallel version of  holonomy corrected LQC in the flat FLRW space-time}

Teleparallelism is a gravitational theory based in the Weitzenb\"ock connection (see for instance
\cite{ha}). It's well-known that General Relativity is equivalent to a teleparallel theory where the Lagrangian is a linear function of the
scalar torsion (see for instance \cite{slb}).
Then, since for the flat FLRW metric  the Lagrangian density is $\frac{1}{2}RV$ which can be written as follows
\begin{eqnarray}
\frac{1}{2}RV=-3H^2V+\ddot{V}.
\end{eqnarray}
where $\ddot{V}$ is a total derivative, we can conclude  that  General Relativity in the flat FLRW geometry can be built with the Lagrangian
$\frac{1}{2}TV$, where
 $T=-6H^2$ is the scalar torsion.

\vspace{0.5cm}

In general,  teleparallel
$F(T)$ gravity  in the flat FLRW geometry
 is based in the Lagrangian density ${\mathcal L}_T=VF(T)+{\mathcal L}_M$,
where ${\mathcal L}_M$ is the matter  Lagrangian density.
After Legendre's transformation, the Langrangian density gives rise to
the following Hamiltonian density
\begin{eqnarray}\label{9}
{\mathcal H}_T= \left(2T\frac{dF(T)}{dT}-F(T)  +\rho \right)V.\end{eqnarray}

Then, the Hamiltonian constrain  ${\mathcal H}_T=0$ leads to
the modified Friedmann equation  (see for instance \cite{aho})
\begin{eqnarray}\label{10}
\rho=-2 \frac{dF(T)}{dT}T+F(T)\equiv G(T),
\end{eqnarray}
which depicts a curve in the plane $(H,\rho)$.

Conversely, given a curve of the form $\rho=G(T)$
for some function $G$, the way to reconstruct the corresponding  Lagrangian density ${\mathcal L}_T$ , consists in
 integrating    the modified Friedmann
equation  obtaining as a result
\begin{eqnarray}\label{11}
F(T)=-\frac{\sqrt{-T}}{2}\int \frac{G(T)}{T\sqrt{-T}}dT.
\end{eqnarray}

Then, the idea of teleparallel LQC is to find an explicit $F(T)$ theory whose modified Friedmann equation coincides with the ellipse (\ref{7}).
This could be done
 splitting the ellipse  in two pieces $\rho=G_-(T)$ (the branch  where $\dot{H}<0$) and $\rho=G_+(T)$ (the branch  where $\dot{H}>0$), with
\begin{eqnarray}\label{8}
G_{\pm}(T)=\frac{\rho_c}{2}\left(1\pm\sqrt{1+\frac{2T}{\rho_c}}  \right),\end{eqnarray}
and $T=-6H^2$ (the scalar torsion in the flat FLRW space-time).

Finally, from formula (\ref{11}), the holonomy corrected Friedmann equation has been recently  obtained in \cite{aho}  using the following function
\begin{eqnarray}\label{12}
 F_{\pm}(T)=\pm\sqrt{-\frac{T\rho_c}{2}}\arcsin\left(\sqrt{-\frac{2T}{\rho_c}}\right)+G_{\pm}(T),
\end{eqnarray}
which is the basis of the teleparallel formulation of LQC.

\section{Scalar cosmological perturbations}

In this Section we  obtain the equations for scalar perturbations in teleparallel LQC and we compare them with
the corresponding ones in holonomy corrected LQC.

For simplicity, we work
 in longitudinal gauge $ds^2=(1+2\Phi)dt^2-a^2(1-2\Phi)d{\bf x}^2$, where
$\Phi$ is the Newtonian potential, and we consider a scalar field
 with Lagrangian density
\begin{equation}\label{13}
 {\mathcal L}_M=\left(\frac{1}{2}\dot{{\varphi}}^2-V({\varphi})\right)V,
\end{equation}
with $\varphi=\bar{\varphi}+\delta{\varphi}$, being $\bar{\varphi}$ the homegeneous part of the field.

The perturbation equations in teleparallel   $F(T)$ gravity are (see for example \cite{ccdds})
\begin{eqnarray}\label{15a}
 2\frac{dF}{dT}\frac{\Delta\Phi}{a^2}+6H\frac{dG}{dT}\dot{\Phi}-T\frac{dG}{dT}{\Phi}=
 \frac{1}{2}\left(\dot{\bar{\varphi}}(\delta\dot{\varphi}-\dot{\bar{\varphi}}\Phi)
+\frac{dV(\bar{\varphi})}{d\varphi}\delta{\varphi}   \right).
\end{eqnarray}
\begin{eqnarray}\label{16a}
 -2\frac{dG}{dT}(\dot{\Phi}+H{\Phi})=\frac{1}{2}\dot{\bar{\varphi}}\delta{\varphi}.
\end{eqnarray}
\begin{eqnarray}\label{17a}&&
 -2\frac{dG}{dT}\ddot{\Phi}+
\left[-8H\frac{dG}{dT}+6\dot{T}\left(\frac{d^2F}{dT^2}
+\frac{2}{3}T\frac{d^3F}{dT^3}\right)\right]\dot{\Phi}+
%\nonumber\\&&
\left[T\frac{dG}{dT}+4\dot{H}\times\right.
\nonumber\\&&\left.
\left(\frac{dF}{dT}+5T\frac{d^2F}{dT^2}+{2}T^2\frac{d^3F}{dT^3}\right)
\right]{\Phi}
=\frac{1}{2}\left(\dot{\bar{\varphi}}(\delta\dot{\varphi}-\dot{\bar{\varphi}}\Phi)
-\frac{dV(\bar{\varphi})}{d\varphi}\delta{\varphi}   \right)
\end{eqnarray}
where $G$ has been introduced in equation (\ref{10}).

To obtain the equations for scalar perturbations in  teleparallel LQC, we have to choose  the $F(T)$ theory defined by equation (\ref{12}) and
inserting it in the general equations (eq's (\ref{15a}), (\ref{16a}) and (\ref{17a})).  The final form of our equations is
\begin{eqnarray}\label{14}
\tilde{c}^2_{s}\Delta \Phi-3{\mathcal H}\Phi'
-\left({\mathcal H}'+2{\mathcal H}^2\right)\Phi=
%\nonumber\\&&
\frac{\Omega}{2}\left(\delta\varphi'\bar{\varphi}'+\delta\varphi
\frac{d V(\bar{\varphi})}{d\varphi}a^2 \right).
\end{eqnarray}
\begin{eqnarray}\label{15}
 \Phi'+{\mathcal H}\Phi=
\frac{\Omega}{2}\delta\varphi\bar{\varphi}'.
\end{eqnarray}
\begin{eqnarray}\label{16}
 \Phi''+\left(3{\mathcal H}-2\epsilon\right)\Phi'
+ \left({\mathcal H}'+2{\mathcal H}^2-2{\mathcal H}\epsilon\right)\Phi=
%\nonumber\\&&
\frac{\Omega}{2}\left(\delta\varphi'\bar{\varphi}'-\delta\varphi
\frac{d V(\bar{\varphi})}{d\varphi}a^2 \right),
\end{eqnarray}
where the following notation has been introduced:
\begin{enumerate}
 \item $\Omega=-\frac{1}{2G_{,T}}=1-\frac{2\rho}{\rho_c}$.
\item $\epsilon=\frac{1}{2}\frac{\Omega'}{\Omega}$
\item The square of the velocity of sound is equal to $\tilde{c}_{s}^2=2|\Omega| \left|\frac{d F_{\pm}(T)}{dT}\right|$, being
\begin{eqnarray}\left|\frac{d F_{\pm}(T)}{dT}\right|=
\frac{\sqrt{\frac{\rho_c}{\rho}}\arcsin\left(2\sqrt{\frac{\rho}{\rho_c}}\sqrt{1-\frac{\rho}{\rho_c}}\right)}{4\sqrt{1-\frac{\rho}{\rho_c}}}.
\end{eqnarray}
\item $'=\frac{d}{d\eta}$ is the derivate with respect to the conformal time $\eta$.
\end{enumerate}

Combining these equations one obtains, in teleparallel LQC, the dynamical equation for the
 Newtonian potential
\begin{eqnarray}\label{holonomy}
 \Phi''-\tilde{c}^2_{s}\Delta \Phi+2\left({\mathcal H}- \left(\frac{{\bar{\varphi}}''}{{\bar{\varphi}}'}+\epsilon\right)\right)\Phi'
%\nonumber\\&&
+2\left({\mathcal H}'- {\mathcal H}\left(\frac{{\bar{\varphi}}''}{{\bar{\varphi}}'}+\epsilon\right)\right)\Phi=0.
\end{eqnarray}

On the other hand  in \cite{cmbg}, using holonomy corrected LQC, the authors  obtained the
same equations (\ref{14})-(\ref{holonomy}),
but with a square of the velocity of sound equal to ${c}_{s}^2=\Omega.$
As a consequence,
in holonomy corrected LQC,
$c_s^2>0$ when $\rho<\rho_c/2$, whereas when  $\rho>\rho_c/2$ one has $c_s^2<0$. The latter means that
in the super-inflationary phase
the holonomy corrected equation that corresponds to  (\ref{holonomy}), i.e. equation (\ref{holonomy}) where $\tilde{c}_s$  is replaced by $c_s$,
 becomes elliptic.
 This behavior never happens
in our teleparallel formulation of LQC,  where
$\tilde{c}^2_{s}$
is always positive and, thus, the equation is always hyperbolic.

\vspace{0.5cm}

The dynamical equation for the perturbed scalar field $\delta\varphi$, also depends on the formulation used. In teleparallel LQC, since
the matter Lagrangian is the same as in standard Einstein Cosmology (EC),  this equation coincides with the usual one,
 that is, it is given by
\begin{eqnarray}\label{18}
 {\delta\varphi}''+2{\mathcal H}{\delta\varphi}'-\Delta{\delta\varphi}+a^2\frac{d^2V(\bar{\varphi})}{d\varphi^2} \delta\varphi
+2a^2\frac{dV(\bar{\varphi})}{d\varphi}\Phi-4 \bar{\varphi}'\Phi'=0.
\end{eqnarray}

However in holonomy corrected LQC, a counter-term is added to the matter Hamiltonian in order to
close the algebra of total constrains, giving rise to an equation
which differs with the standard one,  by a square of the  velocity of sound equal to $c_s^2=\Omega$  (see  \cite{cmbg}).

Here a  remark is in order:
{\it If one does not introduce any counter-term in the matter Hamiltonian, then  $\delta\varphi$ satisfies the standard
equation (\ref{18}). As a consequence, the algebra of constrains ceases to be closed, in the sense that instead of the bracket $$\{{\mathcal H}_{m+g}[N_1],
{\mathcal H}_{m+g}[N_2]\}=\Omega D_{m+g}\left[\frac{1}{\sqrt{\bar{p}}}\partial^{a}(\delta N_2 -\delta N_1)\right],$$
where ${\mathcal H}_{m+g}$ and  $D_{m+g}$ are the total hamiltonian and diffeomorphism constrain \cite{cbvg},
one has
$$\{{\mathcal H}_{m+g}[N_1],
{\mathcal H}_{m+g}[N_2]\}=D_{m}\left[\frac{1}{\sqrt{\bar{p}}}\partial^{a}(\delta N_2 -\delta N_1)\right],$$
where $D_{m}$ is  only the matter part of the diffeomorphism constrain.}

\subsection{Mukhanov-Sasaki equations for scalar perturbations}

Once we have obtained the equations of scalar perturbations, introducing, as in standard cosmology,
(see for example formulae (8.56)-(8.58) of \cite{m})
the
Mukhanov-Sasaki (M-S) variables
\begin{eqnarray}\label{19}
 v=a(\delta\varphi+\frac{\bar{\varphi}'}{{\mathcal H}}\Phi); \quad z=\frac{a\bar{\varphi}'}{{\mathcal H}},
\end{eqnarray}
equation (\ref{holonomy}) with $\tilde{c}_s$  replaced by $c_s$ (the corresponding equation in holonomy corrected LQC) becomes the following  M-S equation
\begin{eqnarray}\label{21}
 v''-c^2_{s}\Delta v-\frac{z''}{z}v=0.
\end{eqnarray}

On the other hand,
in LQC as $F(T)$ gravity, to obtain the corresponding M-S equation, following \cite{m}
we write equations (\ref{14}) and (\ref{15}) as follows
\begin{eqnarray}\label{eq1}
\tilde{c}_s^2\Delta \Phi=\frac{\Omega (\bar{\varphi}')^2}{2{\mathcal H}}\left({\mathcal H}\frac{\delta\varphi}{\bar{\varphi}'}+\Phi\right)',
\end{eqnarray}
and
\begin{eqnarray}\label{eq2}
\left(a^2\frac{\Phi}{{\mathcal H}}\right)'=\frac{\Omega a^2(\bar{\varphi}')^2}{2{\mathcal H}^2}\left({\mathcal H}\frac{\delta\varphi}{\bar{\varphi}'}+\Phi\right).
\end{eqnarray}

Then, introducing the variables
\begin{eqnarray}\label{22}  v=a\frac{\sqrt{|\Omega|}}{\tilde{c}_{s}}(\delta\varphi+\frac{\bar{\varphi}'}{{\mathcal H}}\Phi);\quad
z=\frac{a\sqrt{|\Omega|}\bar{\varphi}'}{\tilde{c}_{s}{\mathcal H}},
\end{eqnarray}
and
\begin{eqnarray}\label{24}
 u=\frac{2a\Phi\sqrt{|\Omega|}}{\bar{\varphi}'\Omega};
\quad\theta=\frac{1}{\tilde{c}_{s}z},
\end{eqnarray}
 equations (\ref{eq1}) and (\ref{eq2}) become
\begin{eqnarray}\label{25}
 \tilde{c}_{s}\Delta u= z\left(\frac{v}{z}\right)';\quad \theta\left(\frac{u}{\theta}\right)'= \tilde{c}_{s}v.
\end{eqnarray}
Performing the Laplacian in the second equation and using the first one, one gets the M-S equation
\begin{eqnarray}\label{26}
 v''-\tilde{c}^2_{s}\Delta v-\frac{z''}{z}v=0.
\end{eqnarray}

Finally, note that, in both formulations,  the variable $v$ is related to the
curvature fluctuation in co-moving coordinates
\begin{eqnarray}\label{27}
 \zeta=\Phi-\frac{{H}}{\dot{ H}}(\dot{\Phi}+{H}\Phi),
\end{eqnarray}
by the relation $v=z\zeta$.

\section{Scalar power spectrum in a matter bounce scenario}

In this Section we will see that, as in holonomy corrected LQC  \cite{w}, for a matter-dominated universe where the scale factor, the Hubble parameter
and the energy density  are given by
\begin{eqnarray}\label{lqc}
a(t)=\left(\frac{3}{4}\rho_ct^2+1\right)^{1/3}, \quad H(t)=\frac{\frac{1}{2}\rho_ct}{\frac{3}{4}\rho_ct^2+1}\quad\mbox{and}\quad
\rho(t)=\frac{\rho_c}{\frac{3}{4}\rho_ct^2+1},
\end{eqnarray}
in teleparallel LQC,
 the scalar power spectrum
is also scale invariant. This is in agreement with the fact that a matter-dominated universe in the contracting phase leads
to a scale invariant spectrum of perturbations \cite{wands,pn}.

When the energy density is small ($\rho\ll \rho_c$), EC is recovered and equation (\ref{26}) becomes the usual M-S
equation that, working in Fourier space, for a matter-dominated universe  is given by
\begin{eqnarray}\label{29}
 v_k''+\left(k^2 -\frac{a''}{a}\right)v_k=0\Leftrightarrow
 v_k''+\left(k^2 -\frac{2}{\eta^2}\right)v_k=0.
\end{eqnarray}

Assuming that at early times the universe is in the Bunch-Davies (adiabatic) vacuum, one must take for $\eta\rightarrow-\infty$
\begin{eqnarray}\label{30}
 v_k(\eta)=\sqrt{\frac{-\pi\eta}{4}}H^{(1)}_{3/2}(-k\eta)=\frac{e^{-ik\eta}}{\sqrt{2k}}\left(1-\frac{i}{k\eta}\right).
\end{eqnarray}

At early times all the modes are inside the Hubble radius, and when time moves forward the modes leave this radius. For a matter-dominated universe
in EC, the modes well outside the Hubble radius are characterized by the condition
\begin{eqnarray}\label{31}
 k^2\eta^2 \ll 1\Longleftrightarrow k^2 \ll \left|\frac{a''}{a}\right|\Longleftrightarrow k^2 \ll \left|\frac{1}{\tilde{c}^2_{s}}\frac{z''}{z}\right|,
\end{eqnarray}
because for small values of $\rho$ one has $z\cong\frac{\sqrt{3}a}{\sqrt{1-\frac{\rho}{\rho_c}}}\cong\sqrt{3}a.$

Then, when holonomy effects are not important,  for modes well outside the Hubble radius the M-S equation becomes
\begin{eqnarray}\label{32}
 v_k'' -\frac{z''}{z}v_k=0,
\end{eqnarray}
which can be solved using the method of reduction of the order, giving as a result
\begin{eqnarray}\label{33}
 v_k(\eta)=B_1(k)z(\eta)+B_2(k)z(\eta)\int_{-\infty}^{\eta}\frac{d\bar{\eta}}{z^2(\bar{\eta})},
\end{eqnarray}
which means that
 at early times in the contracting phase, for modes well outside the Hubble radius, the expressions (\ref{30}) and (\ref{33}) give the same
solution. The solution given by (\ref{30}) could be expanded in terms of $k\eta\ll 1$, and retaining  the leading terms in the real and imaginary parts of $v_k$,
one gets
\begin{eqnarray}
v_k(\eta)\cong -\frac{k^{3/2}\eta^2}{3\sqrt{2}}-\frac{i}{\sqrt{2}k^{3/2}\eta}.
\end{eqnarray}

On the other hand,
the explicit solution of
(\ref{33}) is obtained using the approximation  $z\cong\sqrt{3}a=\frac{1}{4\sqrt{3}}\rho_c\eta^2$,
where we have used the well-known classical relation $a=\frac{\rho_c}{12}\eta^2$
obtained  performing the approximation $a(t)\cong \left(\frac{3\rho_c}{4} \right)^{1/3}t^{2/3}$ in (\ref{lqc})
and using the relation between conformal and cosmological time $\eta=\int \frac{1}{a(t)}dt$
 (see for instance \cite{w}).
Then formula (\ref{33}) gives as a result
\begin{eqnarray}
v_k(\eta)\cong\frac{B_1(k)}{4\sqrt{3}}\rho_c\eta^2-\frac{4B_2(k)}{\sqrt{3}\rho_c}\frac{1}{\eta}.
\end{eqnarray}

 Matching both solutions one obtains
 \begin{eqnarray}\label{34}
 B_1(k)= \sqrt{\frac{8}{3}}\frac{2k^{3/2}}{\rho_c}  \quad \mbox{and}\quad B_2(k)=i\sqrt{\frac{3}{8}}\frac{\rho_c}{2k^{3/2}}.
\end{eqnarray}

Once we have calculated the coefficients $B_1(k)$ and $B_2(k)$ we use equation (\ref{33}) to calculate $v_k$ at late times. More precisely,
we calculate   $v_k$  in the classical regime of the expanding phase for modes that are still  well outside of the Hubble radius. Note that we are considering modes that
in the contracting phase leave the Hubble radius and then evolve
satisfying $ k^2 \ll \left|\frac{1}{\tilde{c}^2_{s}}\frac{z''}{z}\right|$. Then, we can approximate $v_k$ by
\begin{eqnarray}\label{35}
 v_k(\eta)=(B_1(k)+B_2(k)R)z(\eta),
\end{eqnarray}
where $R\cong\int_{-\infty}^{\infty}\frac{d\bar{\eta}}{z^2(\bar{\eta})}$, because $\eta$ is large enough.

From (\ref{35}) one has
\begin{eqnarray}\label{36}
 \zeta_k(\eta)=\frac{ v_k(\eta)}{z(\eta)}= B_1(k)+B_2(k)R\cong B_2(k)R ,
\end{eqnarray}
and thus, the scalar power spectrum is given by
\begin{eqnarray}\label{37}
 {\mathcal P}_{\zeta}(k) \equiv
 \frac{k^3}{2\pi^2}|\zeta_k(\eta)|^2
 =\frac{ 3\rho_c^2}{64\pi^2}R^2
%\nonumber\\&&
 =\frac{\rho_c}{144\pi^2}\left(
\int_0^{\pi/2}\frac{x}{\sin x}dx \right)^2
= \frac{\rho_c}{36\pi^2}{\mathcal C }^2,
\end{eqnarray}
where ${\mathcal C}=1-\frac{1}{3^2}+\frac{1}{5^2}-\frac{1}{7^2}+...= 0.915965...$ is Catalan's constant
 (see formula $3.747 (2)$ of \cite{gr}).

On the other hand,
in holonomy corrected LQC  $z(t)= \frac{2a^{5/2}(t)}{\sqrt{\rho_c}|t|}$ which leads to a simple calculation of
$R^2$ giving as a result $\frac{\pi^2}{27\rho_c}$. Consequently in holonomy corrected LQC one has (see \cite{w})
\begin{eqnarray}\label{38}
 {\mathcal P}_{\zeta}(k)=\frac{\rho_c}{576}.
\end{eqnarray}

Formulas (\ref{37}) and  (\ref{38}) have very important consequences, because as has been pointed out in \cite{w},
since $\rho_{pl}=64\pi^2$,
to agree with the observed value ${\mathcal P}_{\zeta}(k)\simeq 2\times 10^{-9}$
 \cite{btw}
 one has to take $\rho_c\sim 10^{-9}\rho_{pl}$, which contradicts  its current value $\rho_c\cong 0.4\rho_{pl}$.
In fact, since $\rho_c=\frac{2\sqrt{3}}{\gamma^3}$, to get  $\rho_c  \sim 10^{-9}\rho_{pl}$, one has to take as a  value of the Barbero-Immirzi
parameter $\gamma\sim 10^{2}$, which is greater than its current value $0.2375$ obtained relating the black hole entropy in LQC with
the Bekenstein-Hawking entropy formula \cite{meissner}.
This contradiction does not appear in our version of LQC, where $\rho_c$ is understood as a parameter,
whose value
has to be obtained from  observations,
 seems to be
close to  $10^{-9}\rho_{pl}$, and thus, for this  critical density,  geometric quantum
effects do not appear to
affect the evolution of the universe.

Note also that, the value of $\rho_c$ is smaller that $\rho_{pl}$ but is still  two orders greater than
the most natural value of the initial energy density  in chaotic inflation. This initial energy density
 could be deduced
as follows: In inflationary cosmology one has the general formula for the spectrum of scalar perturbations \cite{btw}
\begin{eqnarray}
 {\mathcal P}_{\zeta}(k)=\frac{V^3(\bar{\varphi}_i)}{12\left(\frac{d V(\bar{\varphi}_i)}{d\varphi}\right)^2},
\end{eqnarray}
where $\bar{\varphi}_i$ is the value of the field at the beginning of infation.

Applying this formula, for example, to the quadratic potential $V(\varphi)=\frac{1}{2}m^2\varphi^2$ one gets
\begin{eqnarray}\label{inflation}
 {\mathcal P}_{\zeta}(k)=\frac{m^2\bar{\varphi}_i^4}{96\pi^2}=\frac{4\rho_i}{3\rho_{pl}}\bar{\varphi}_i^2,
\end{eqnarray}
where $\rho_i$ is the initial energy density.

On the other hand, at the end of inflation the scalar factor is given by \cite{fmsvv}
\begin{eqnarray}
 a_e=a_ie^{\frac{1}{4}(\bar{\varphi}_i^2-\bar{\varphi}_e^2)}\cong a_ie^{\frac{1}{4}\bar{\varphi}_i^2},
\end{eqnarray}
where $a_i$ is the value of the scalar factor at the beginning of inflation and $\bar{\varphi}_e$ is the value of the field at the end of
inflationary phase.

Assuming that inflation produces an expansion of $60$ e-folds
(needed in order to solve the horizon and flatness problems in EC) one obtains $\bar{\varphi}_i^2=240$. Then, inserting this value in
(\ref{inflation}) and using
 the constrain ${\mathcal P}_{\zeta}(k)\simeq 2\times 10^{-9}$, one concludes that $\rho_i\sim 10^{-11}\rho_{pl}$.

To end this Section two minor remarks are in order:
{\it \begin{enumerate}\item
The key point to obtain the scale invariant power spectrum  (\ref{37}) (resp. (\ref{38})) is that one only considers
modes that after leaving and before re-entering the Hubble radius satisfy $k^2 \ll \left|\frac{1}{\tilde{c}^2_{s}}\frac{z''}{z}\right|$
(resp. $k^2 \ll \left|\frac{1}{c^2_{s}}\frac{z''}{z}\right|$), that is, the term $\tilde{c}^2_{s}\Delta v$ (resp. $c^2_{s}\Delta v$) in the M-S
equation is disregarded between the leaving and the reentry of the modes in the  Hubble radius.
\item The same kind of calculation could be done for a universe with equation of state $P=\omega \rho$ with $|\omega|\ll 1$.
The calculation of the power spectrum is
more involved, but  the spectral index $n_{\zeta}$ could be easily calculated from the  dominant term in the  asymptotic expression of the
corresponding
Bunch-Davies  vacuum state
 $v_k(\eta)\cong\sqrt{\frac{-\pi\eta}{4}}H^{(1)}_{3/2-6\omega}(-k\eta)\sim k^{-3/2+6\omega}$, giving as a result
\begin{eqnarray}
 n_{\zeta}\equiv 1+\frac{\ln d{\mathcal P}_{\zeta}(k)}{d\ln k}=1+12\omega.
\end{eqnarray}

\end{enumerate}
}

\section{Tensor cosmological perturbations}

In teleparallel LQC, the equation of perturbations can be obtained inserting (\ref{12}) in  the general equation \cite{cdds}
\begin{eqnarray}\label{39}
\frac {dF(T)}{dT}\left(\ddot{h}_i^a-\frac{\Delta{h}_i^a }{a^2}+3H\dot{h}_i^a  \right)+{\dot{T}\dot{h}_i^a }\frac {d^2F(T)}{dT^2}
=0.
\end{eqnarray}

Performing the change of variables
\begin{eqnarray}\label{40}
z_t\equiv \frac{a\tilde{c}_s}{\sqrt{2|\Omega|}},\quad v_t\equiv hz_t,
\end{eqnarray}
where $h$ represents the two degrees of freedom of $h_i^a$, we have obtained the following M-S equation for tensor perturbations
\begin{eqnarray}\label{41}
 v_t''-\Delta v_t-\frac{z_t''}{z_t}v_t=0.
\end{eqnarray}

On the other hand, using holonomy corrections without counter-terms,  in \cite{bh} the authors obtained
\begin{equation}\label{42}
 {h_i^{a }}''+2{\mathcal H}{h_i^{a }}'
-\Delta {h_i^{a }}+ {\mathcal H}^2\frac{\rho}{\rho_c-\rho}{h_i^{a }}=0,
\end{equation}
that after the change of variables $v_t=ah$  becomes
\begin{eqnarray}\label{43}
 v_t''-\Delta v_t-\left(\frac{a''}{a}-{\mathcal H}^2\frac{\rho}{\rho_c-\rho} \right)v_t=0,
\end{eqnarray}
which has, as equation (\ref{41}), a velocity of sound equal to $1$, but does not have the standard form of a M-S equation.

However, in \cite{cbvg} (see also \cite{clb}), to avoid anomalies, as in the case of scalar perturbations,
the authors use counter-terms, obtaining an equation for tensor perturbations of the form
\begin{equation}\label{45}
 {h_i^{a }}''+2\left({\mathcal H}-\epsilon\right){h_i^{a }}'
-\Omega\Delta {h_i^{a }}=0,
\end{equation}
which after the change of variables  $z_t\equiv {a }{{\Omega}}^{-1/2}$ and $v_t\equiv h z_t$ becomes
 the following M-S equation
\begin{eqnarray}\label{44}
 v_t''-c^2_s\Delta v_t-\frac{z_t''}{z_t}v_t=0.
\end{eqnarray}

It is important to realize that
in the
super-inflationary phase ($\rho>\rho_c/2$), the variable $z_t$ becomes imaginary, which  is a great difference with the classical M-S equation
where $z_t=a$. Moreover,
equation (\ref{44}) has two singular points at the beginning and  end of the super-inflationary phase,
when $\rho=\rho_c/2$,
which means that there is not any objective criterium of continuity to define the solution at these points, and thus, there are
infinite ways to match solutions at these points. Consequently, infinite mode functions could be used to calculate the power spectrum
of tensor perturbations. For example,
 when holonomy corrections are taken into account,
  for the modes we are considering, $z_t=
{a }{{\Omega}}^{-1/2}$ is a solution, but $\tilde{z}_t={a }{{|\Omega}|}^{-1/2}$ is another one.
In fact,  one can build
infinite solutions, because we cannot impose any kind continuity  at the singular points.

Using $z_t$ as a solution and following the same steps as in the case of scalar perturbations,
 the ratio of tensor to scalar  perturbations for a universe with
equation of state $P=\omega \rho$, will be $r\equiv\frac{{\mathcal P}_{h}(k)}{{\mathcal P}_{\zeta}(k)}\sim \omega^2$.
Since the observed scalar index $n_{\zeta}=1+12\omega$ is approximately $0.968$ one concludes that $\omega\cong -3 \times 10^{-3}$, which
shows that,
like in the slow-roll phase of inflation,
the amplitude of tensor perturbations is suppressed relative to that of scalar ones \cite{w}.
However, using $\tilde{z}_t$, for a matter-dominated universe, we have  obtained $r\cong 27/\pi^2 $ which is of  order $1$.

On the other hand, due to the different definitions of $z_t$ in both formulations,
our M-S equation for tensor perturbations
does not have any singular point. The solution $z_t=\frac{a\tilde{c}_s}{\sqrt{2|\Omega|}}$ is unambiguously defined,
giving a ratio of tensor to scalar perturbations  of order $1$, in agreement with the current calculations in $F(T)$ gravity (see for instance
\cite{ccdds}).

However  the
current CMB bound is $r\lesssim 0.2$.
Consequently,  a mechanism to amplify the scalar perturbations which could be done,
as in quintom bounce scenario for the Lee-Wick theory \cite{cbz},
 by introducing
a curvaton field, has to be considered.
Another recent  alternative (see, for instance
\cite{ceb}) to amplify scalar perturbations in the framework of EC, which is free of ghost fiels, consists on introducing
higher order operators in the matter Lagrangian.
 In our case of teleparallel LQC, these possible solutions  deserve future investigations.

\subsection{A possible viable model}

Here, we will suggest another alternative solution to this problem.
The ratio of tensor to scalar perturbations in LQC, for a matter-dominated universe, is approximately
\begin{eqnarray}
r\cong \frac{1}{6}\frac{\left(\int_{-\infty}^{\infty}\frac{1}{z_t^{2}}d\eta\right)^2}{\left(\int_{-\infty}^{\infty}\frac{1}{z^{2}}d\eta\right)^2},
\end{eqnarray}
where $z$ and $z_t$ were introduced in equations (\ref{22}) and (\ref{40}).
In this expression, both the numerator and the denominator are of the order $\frac{1}{{\rho_c}}$, and thus, the ratio $r$ is of the order $1$.

On the other hand, the variables $z$ and $z_t$ are related by
\begin{eqnarray}
\frac{1}{z^2}=\frac{1}{z_t^2}\frac{H^2}{|\dot{H}|}\frac{\tilde{c}_s^4}{|\Omega|}.
\end{eqnarray}

Since $\frac{\dot{H}}{H^2}$ is an slow roll parameter,
the idea is that there is a phase transition from a matter-dominated to a quasi de Sitter stage. Then, when the universe is in this quasi
de Sitter phase one has $\frac{1}{z^2}\gg \frac{1}{z_t^2}$, and thus, $r$ will increase. Note that this stage cannot happen when the
universe bounces  ($\rho=\rho_c$) or when it enters in the super-inflationary phase ($\rho=\rho_c/2$), because $\frac{1}{z^2}$ vanishes at
these energies.

For a sake of simplicity, we choose an abrupt phase transition in the contracting phase, at energy density $\rho=\rho_c/4$. This
happens at time $t_i=-\frac{2}{\sqrt{\rho_c}}$, and we consider the following dynamics during, for example, the period of
time between $-\frac{2}{\sqrt{\rho_c}}$ and $-\frac{1}{\sqrt{\rho_c}}$
\begin{eqnarray}
 H(t)\cong -\frac{\sqrt{\rho_c}}{4}-\sqrt{\epsilon}\rho_c(t-t_i); \quad \rho(t)\cong \rho_c/4,
\end{eqnarray}
where $\epsilon\ll 1$ is a dimensionless parameter. In that case the scale factor evolves as follows
\begin{eqnarray} a(t)\cong 4^{1/3}e^{\sqrt{\rho_c}(t_i-t)/4},\end{eqnarray}
and
  a simple calculation yields the following bound
\begin{eqnarray}
 r\sim \epsilon\ll 1.
\end{eqnarray}

 It is well known that, for homogeneous and isotropic cosmologies,  given the value of $H(t)$ for all time, there always exists a potential for
scalar fields
$V(\bar{\varphi})$ which provides the desired  dynamics (see for instance \cite{aho}).
However, this potential will be  somewhat complicated which is, from a physical viewpoint, an unpleasant feature of our possible solution.
In fact, in LQC, for a matter-dominated universe, the potential is given by
\begin{eqnarray}
 V(\bar{\varphi})=\frac{8}{3}\frac{e^{-\sqrt{3}\bar{\varphi}}}{\left(1+\frac{4}{3\rho_c}e^{-\sqrt{3}\bar{\varphi}}\right)^2},
\end{eqnarray}
which has to be flattened at some energy level in order to obtain a quasi de Sitter phase. Moreover, in the expanding phase the potential
must have a local minimum that allows the oscillations of the scalar field in order to decay  creating  light particles,
which finally thermalize  yielding
a hot Friedmann universe that matches with the Standard Model.

Of course, this potential would have a very complicated shape meaning that our solution is somewhat artificial.
To obtain a more convincing solution one also might introduce barotropic fluids. In that case, one may image, at early times, a universe
in the contracting phase   dominated
by a dust fluid (the scalar field would be in the minim of the potential), evolving to a radiation dominated one.
During these periods, since the
universe is contracting, the
field climbs up the potential and will eventually dominate, and thus  a phase transition to the quasi de Sitter phase where scalar perturbation
 amplify would happen. After the universe bounces, in the expanding phase, the scalar field would go down to the minimum
 of the potential
 and the  universe would be radiation
 dominated obtaining a hot Friedmann universe.

\section{Conclusions}

In this paper we have shown that teleparallel LQC could be a viable alternative to inflationary cosmology. The theory mixes properties
of LQC and $F(T)$ gravity, providing a simple non-singular bounce and perturbation equations without singular points (the M-S equation for tensor
perturbations is regular and the curvature fluctuation, in Fourier space, evolves as $\zeta_k(\eta)\sim \int_{-\infty}^{\eta}\frac{1}{z^2(\bar{\eta})}d\bar{\eta}$
which is a regular function). Moreover, if at early times the universe is in a matter-dominated phase (or close to it) the power spectrum of scalar and tensor perturbation will be scale invariant (or nearly scale  invariant). The problem
of our formulation, like
in the other $F(T)$ theories, when one considers a simple matter bounce scenario where the universe is matter-dominated or close to it all
the time, is that the ratio of tensor to scalar perturbations is of order one, in contradiction with the current bound.
This problem could be sort out incorporating to the theory  some complicated
mechanisms (curvaton fields, extra higher order terms in the matter lagrangian density, artificial potentials,...). However
what would be
desirable is to find a simple bouncing scenario which agrees with current observations, which for the moment  does not exist in the current literature.
Here, at the end of the paper we have outlined a  possible solution  which seems easier than the current ones. It is based on mixing a scalar field with some barotropic fluids (dust and radiation) and incorporating  a phase transition from the matter dominated universe to a quasi de Sitter phase in order to
achieve the bound $r\lesssim 0.2$, where $r$ is the ratio of tensor to scalar perturbations.

\vspace{0.5cm}

The author  thanks E. Saridakis and J. Amor\'os for helpful discussions.
This investigation has been supported in part by MINECO (Spain)
MTM2011-27739-C04-01, and by AGAUR (Generalitat de
Catalunya), Contract No. 2009SGR-345.

\end{document}